\documentclass[prl,twocolumn,twoside,showpacs,superscriptaddress,floatfix]{revtex4}
\usepackage{graphics,amssymb,amsmath}

\begin{document}
%\title{New perspectives on collective synchronization}
\title{Collective phase synchronization in locally-coupled limit-cycle oscillators}

\author{H. Hong}
\affiliation{School of Physics, Korea Institute for Advanced Study, 
Seoul 130-722, Korea}

\author{Hyunggyu Park}
\affiliation{School of Physics, Korea Institute for Advanced Study, 
Seoul 130-722, Korea}

\author{M. Y. Choi}
\affiliation{Department of Physics, Seoul National University, Seoul 151-747, Korea}
\affiliation{School of Physics, Korea Institute for Advanced Study, 
Seoul 130-722, Korea}

\date{\today}

\begin{abstract}
We study collective behavior of locally-coupled limit-cycle oscillators 
with scattered intrinsic frequencies on $d$-dimensional lattices.  A linear 
analysis shows that the system should be 
always desynchronized up to $d=4$.  On the other hand, numerical investigation 
for $d= 5$ and 6 reveals the emergence of the synchronized (ordered) phase via a
continuous transition from the fully random desynchronized phase. 
%Its transition nature is not mean-field-like.
This demonstrates that the lower critical dimension for the phase 
synchronization in this system is $d_{l}=4$.
% and its upper critical dimension $d_{uc}>6$. 
\end{abstract}
\pacs{05.45.Xt, 89.75.-k, 05.10.-a}

\maketitle

Up to date, much attention has been paid to the collective behavior of 
coupled nonlinear oscillators since those systems of oscillators
have been known to exhibit remarkable phenomena of synchronization~\cite{ref:synch}.
Those phenomena have been observed in a number of physical, biological, and 
chemical systems, and understood rather well in terms of various model systems. 
In the system of globally-coupled oscillators such as 
the Kuramoto model~\cite{ref:Kuramoto,ref:Tanaka},
the mean-field (MF) theory is valid and yields analytic results to unveil the 
phase transition~\cite{ref:Kuramoto}.  
Systems of locally-coupled oscillators, on the other hand, have not 
been much studied even though local coupling in the system is more 
realistic in nature. 
In some existing studies~\cite{ref:Sakaguchi,ref:Strogatz,ref:Aoyagi} 
collective synchronization, in particular, frequency entrainment has 
been investigated.  However, even numerical results as well as the 
analytic ones including heuristic arguments do not provide a clear 
answer about the question of the lower critical dimension for the 
frequency entrainment.  
Phase synchronization has been also studied in the previous 
studies~\cite{ref:Pikovsky}, however, there are still many fundamental 
questions that are not answered.

In this Letter, we consider a system of locally-coupled oscillators on 
$d$-dimensional lattices and use the relation with a typical model of growing surface,
which allows a linear analysis to show the absence of synchronization
up to $d=4$.  On the other hand, numerical investigation performed
for $d= 5$ and 6 reveals the emergence of the synchronized (ordered) phase via a
continuous transition, indicating the lower critical dimension for 
phase synchronization $d_{l}=4$.

We begin with the set of equations of motion governing the dynamics of
$N$ coupled oscillators located at sites of a $d$-dimensional hypercubic
lattice:
\begin{equation}
\frac{d\phi_i}{dt} = \omega_i - K\sum_{j\in \Lambda_i}\sin(\phi_i - \phi_j),
\label{eq:model}
\end{equation}
where $\phi_i$ and $\omega_i$ stand for the phase and the intrinsic frequency 
of the $i$th oscillator $(i=1,2,...,N)$, respectively.
The intrinsic frequencies are assumed to be randomly distributed 
according to the Gaussian distribution function $g(\omega)$ with
mean $\omega_0$ and variance $2\sigma$. 
For simplicity, we set  $\omega_0 \equiv 0$ without loss of generality. 
The second term on the right-hand side 
represents local interactions between the $i$th oscillator 
and its nearest neighbors the set of which is denoted by $\Lambda_i$.

Without any interaction ($K=0$), each oscillator evolves with its own intrinsic 
frequency, resulting in that the system becomes trivially desynchronized.  
For $K>0$, the coupling term favors locally ordered (synchronized) states and competes
against the randomizing force due to scattered intrinsic frequencies. 
When the coupling is strong enough to create globally ordered states,
the system should exhibit collective synchronization. 
We here focus on phase synchronization which may be probed by 
the conventional phase order parameter 
\begin{equation}
\Delta\equiv \left\langle \frac{1}{N}\left|\sum_{j=1}^N  e^{i \phi_j}\right|\right\rangle,
\label{eq:order}
\end{equation}
where $\langle\cdots\rangle$ denotes the average over realizations of 
intrinsic frequencies. 
Phase synchronization is then identified by nonzero $\Delta$ in the thermodynamic
limit. 

Analytic results are available at the MF level. 
Namely, in the case of globally coupled oscillators, where each oscillator is coupled with
every other one with equal strength $K/N$, it is known that 
phase synchronization emerges as 
$\Delta\sim (K-K_c)^{\beta}$ with $\beta=1/2$  near the critical coupling
strength $K_c=2/\pi g(0)$~\cite{ref:Kuramoto} while
the correlation length diverges as $\xi\sim |K-K_c|^{-\nu}$ with 
$\nu=1/2$~\cite{ref:HHong}. 

When the oscillators are locally coupled, the system has been little investigated.
%In particular, the lower or upper critical dimension as well as the 
%transition nature for phase synchronization has not been addressed.
%In this Letter, we have devoted our attention to obtaining the lower
%critical dimension and the phase transition nature.
% 
Since the nonlinear nature of the sine coupling term in Eq. (\ref{eq:model})
is the major obstacle toward analytic treatment,
we first suppose that, for sufficiently strong coupling strength $K$, the phase difference 
between any nearest neighboring oscillators is small enough to allow the
expansion of the sine function in the linear regime.  
With the appropriate continuum limit taken in space, 
the linearized evolution equation for the phase $\phi({\bf x},t)$ reads
\begin{equation}
\frac{\partial\phi}{\partial t} = \omega({\bf x}) + K\nabla^2 \phi +
{\cal {O}}(\nabla^4 \phi ),
\label{eq:linear}
\end{equation}
where $\omega ({\bf x})$ are uncorrelated random variables, 
satisfying $\langle\omega({\bf x})\rangle=0$ and 
$\langle\omega({\bf x})\omega({\bf x^\prime})\rangle
= 2\sigma\delta({\bf x}-{\bf x^\prime})$~\cite{explanation1}.
We also relax the constraint $0 \le \phi < 2\pi$ and extend the range of $\phi$ to 
$(-\infty , \infty)$, for convenience. 

With the irrelevant high order terms neglected, this equation reminds us of 
the celebrated Edwards-Wilkinson (EW) equation~\cite{EW}, 
traditionally describing certain surface evolution, by
interpreting the phase $\phi({\bf x},t)$ as the front height of the growing surface.
Note, however, that the noise $\omega({\bf x})$ is generated not by conventional
spatio-temporal disorder but by so-called columnar disorder (with spatial dependence only).
%%Accordingly, one can expect that the dynamic behavior of $N$ locally-coupled oscillators 
%%in the strong-coupling regime should be described by the EW surface growth equation with
%%columnar noise. 

In the context of surface growth models, a central quantity of interest is the surface
fluctuation width $W$ defined by
\begin{equation}
W^2(t) = \frac{1}{L^d}\int^{L} d^d {\bf{x}}
\left\langle \left[ \phi({\bf x},t)- {\bar \phi}(t) \right]^2 \right\rangle,
\label{eq:W}
\end{equation}
where $L$ is the linear size of the $d$-dimensional lattice 
$(L^d = N)$ and ${\bar \phi}(t)$ the spatial average of the phase $\phi({\bf x},t)$. 
By means of the Fourier transforms, one can easily solve Eq.~(\ref{eq:linear}) 
to find in the long time limit $(Kt\gg L^2)$ that the steady-state surface width 
scales for large $L$~\cite{ref:HHong}
\begin{eqnarray} 
\label{eq:W2_result}
W^2  &\sim& ( 2\sigma/ K^2) L^{4-d},  ~~~~~~d <4 \nonumber\\
         &\simeq& (\sigma/4\pi^2 K^2) \ln L ,~~~~d =4 \\
       &\sim&   2\sigma/K^2 ,     ~~~~~~~~~~~~~~~d>4. \nonumber
\end{eqnarray}
At any finite values of $K$, the surface width $W$ thus diverges as $L\rightarrow \infty$ 
for $d\le 4$ whereas it remains finite for $d> 4$.  This indicates that the surface 
is always rough (except at $K=\infty$) for $d\le 4$ and
always smooth (except at $K=0$) for $d>4$. 

It is also straightforward to derive the steady-state probability 
distribution~\cite{ref:HHong}:
\begin{equation}
\label{eq:gaussian}
P[\{\phi\}]\sim \exp \left[-(K^2/4\sigma)\int (\nabla^2\phi)^2d^d {\bf x} \right].
\end{equation}
Notice that the Gaussian property of the probability distribution 
links $W$ analytically to the phase order parameter via $\Delta=\exp [-W^2/2]$. 
Therefore our results for $W$, translated into the phase synchronization language,
show that the oscillators are always desynchronized ($\Delta=0$) for $d\le 4$ 
and always synchronized $(\Delta\neq 0$) for $d>4$ in this linearized model.

Our linear theory is valid in the strong-coupling regime; as the weak coupling regime
is approached, the original (nonlinear) system should be more disordered than the 
prediction of the linear theory. 
This establishes that the full nonlinear system described by Eq. (\ref{eq:model})
should also be desynchronized for $d\le 4$ at any finite $K$. 
For $d>4$, it is reasonable to expect a phase synchronization (roughening) transition 
at a finite value of $K$, although one may not exclude the possibility of 
either the full destruction of the synchronized phase at any finite $K$ or 
the absence of the desynchronized phase at any nonzero $K$. 

Before investigating the full nonlinear system described by Eq. (\ref{eq:model}), 
we consider another standard quantity in surface growth models, the height-height 
correlation function 
$C({\bf{x}},t) \equiv \langle\left[\phi({\bf{x}},t)-\phi({\bf{0}},t)\right]^2\rangle$. 
In the linearized regime governed by Eq. (\ref{eq:linear}), we find the 
steady-state behavior for small $x\equiv |{\bf{x}}|$~\cite{ref:HHong}
\begin{eqnarray} 
\label{eq:C_result}
C(x)  &\sim& ( 2\sigma/ K^2) x^2 L^{2-d},  ~~~~~~d <2 \nonumber\\
      &\simeq& (\sigma/2\pi K^2) x^2 \ln L ,~~~~~~d =2 \\
      &\sim&   (2\sigma/ K^2) x^{4-d} ,     ~~~~~~~~~~d>2. \nonumber
\end{eqnarray} 
Note that for $d\le 2$ the correlation $C(x)$ diverges with system size $L$, 
which implies that the average nearest neighbor phase (height) difference
$G=\langle (\nabla\phi)^2\rangle^{1/2}$ is unbounded for any finite $K$ 
in the thermodynamic limit.  As our linear theory is
based on the boundedness of $|\nabla\phi|$, there is no range of $K$ where
the linear theory applies for $d\le 2$.  In contrast, for $d> 2$, $G$ is finite 
and the linear theory is self-consistent at least for large $K$ where
%$G(K)\lesssim O(1)$.
$G(K)\lesssim {\cal {O}}(1)$.
We now examine the nonlinear effects due to the sine coupling in Eq. (\ref{eq:model}).
Unlike in the linearized case, phase $\phi$ may not be bounded even in a
finite system but diverge eventually with a finite angular velocity, 
once its intrinsic-frequency term wins over the nearest-neighbor coupling term. 
In the weak-coupling regime (for small $K$),
these {\em runaway} oscillators with scattered angular velocities
dominate and their phases become completely random to one another, 
leading to the behavior $\Delta\sim N^{-1/2}=L^{-d/2}$.
On the other hand, in the strong coupling regime where the linear theory applies, 
$\Delta$ vanishes exponentially for $d=3$ and algebraically for $d=4$, 
with an exponent depending on $K$ [see Eq. (\ref{eq:W2_result})].

We integrate numerically Eq. (\ref{eq:model}) and measure the phase 
order parameter at various values of $K$ and $L$ for $d=2$ to $6$. 
For convenience, periodic boundary conditions have been employed and $2\sigma$
has been set equal to unity. 
We start from the uniform initial condition ($\phi_i=0$) for a given set of 
$\{\omega_i\}$, chosen randomly according to the Gaussian distribution 
$g(\omega)\sim \exp (-\omega^2 / 4\sigma )$,
and measure the order parameter $\Delta$ averaged over the data in the steady state,
reached after appropriate transient time $(Kt \gg L^2)$.
Here we have used Heun's method~\cite{Heun} to integrate up 
to $4\times 10^4$ time steps, with the time step $\delta t = 0.05$,
and also average over 100 independent sets of $\{\omega_i\}$.
\begin{figure}
\centering{\resizebox*{!}{10.2cm}{\includegraphics{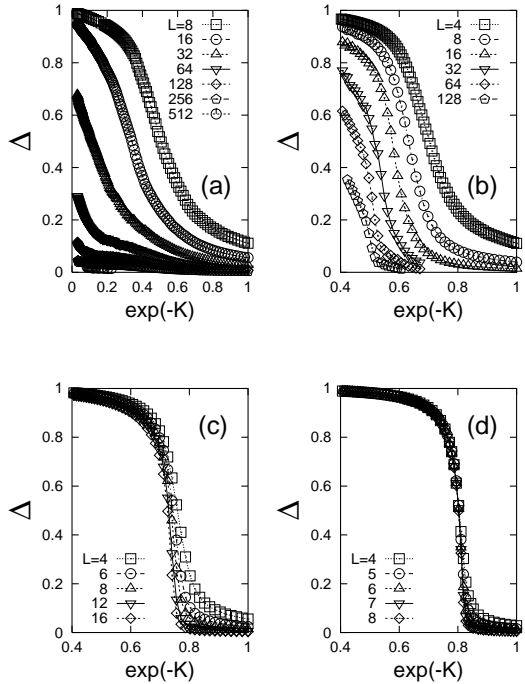}}}
\caption{Behavior of the order parameter $\Delta$ with the coupling strength $K$, 
plotted in terms of $\exp(-K)$, in systems of various size $L$ for (a) $d=2$; 
(b) $d=3$; (c) $d=4$; and (d) $d=5$.  Symbol sizes correspond to 
statistical errors of the data.}
\label{fig:ph_2345D}
\end{figure}
Figure~\ref{fig:ph_2345D} displays the numerical results for the order parameter.
For $d=2$ and $3$, it is clearly observed that the order parameter
decreases rapidly with the system size and seemingly approaches 
zero in the thermodynamic limit for any finite $K$. 
Detailed finite-size analysis~\cite{ref:HHong} shows 
$\Delta\sim L^{-d/2}$ in the weak coupling regime, implying
that phases are completely random and the system is dominated
by runaway oscillators. 
For $d=2$, this fully random phase extends to the regime of large $K$,
while for $d=3$ the linear theory predicting correlated phases 
[see Eq. (\ref{eq:C_result})] 
appears to work for large $K$, 
namely, the data fit well to $\Delta\sim \exp[ -(\sigma/4\pi^3 K^2)L]$
%for $K>K_0$ with $G(K_0)\approx O(1)$.  Numerically, we find that 
for $K>K_0$ with $G(K_0)\approx {\cal {O}}(1)$.  Numerically, we find that 
$K_0\approx \sqrt{2\sigma/\pi}$~\cite{explanation2}.
\begin{figure}
%\centering{\resizebox*{!}{4.5cm}{\includegraphics{4D_ph_L_Keff_1.eps}}}
\centering{\resizebox*{!}{4.5cm}{\includegraphics{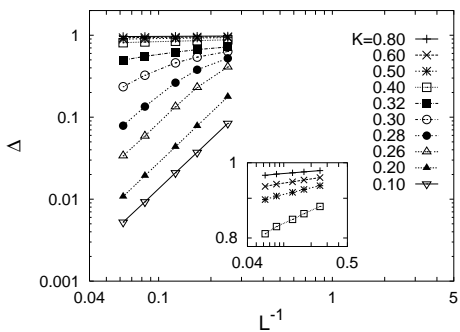}}}
\caption{
Log-log plots of $\Delta$ versus $L^{-1}$ for $d=4$ at various values of 
$K$.  The data for large $K$ are shown in the inset for better visibility. 
Lines are merely guides to eyes.
}
\label{fig:4D_Keff}
\end{figure}
The data for $d=4$ seem to suggest that for large $K$ $\Delta$ remains finite
even in the thermodynamic limit, which contradicts our prediction
based on the linear analysis.  To resolve this puzzle,
we analyze our data carefully by means of finite-size scaling,
and show in Fig.~\ref{fig:4D_Keff} the log-log plots of
$\Delta$ versus $L^{-1}$ at various values of $K$. 
Manifested for $K\lesssim 0.28$ is the fully random phase: $\Delta\sim L^{-2}$.  
For $K\gtrsim 0.40$, $\Delta$ still decreases algebraically with $L$ 
(see the inset of Fig.~\ref{fig:4D_Keff}): $\Delta\sim L^{-\delta(K)}$. 
%albeit with smaller exponents. 
It is pleasing that our data for $K\gtrsim 0.40$ agree perfectly with 
the prediction of the linear theory, $\delta(K)=\sigma/8\pi^2K^2$ 
from Eq. (\ref{eq:W2_result}). 
This result confirms that there is no synchronized phase at any finite $K$
for $d=4$.  It would be interesting to explore the possibility of a phase transition 
near $K\approx K_0= \sqrt{\sigma/4}$ between the fully random phase and 
the critical phase described by the linear theory; this is currently under investigation. 
\begin{figure}
%\centering{\resizebox*{!}{4.5cm}{\includegraphics{5D_ph_L.eps}}}
\centering{\resizebox*{!}{4.5cm}{\includegraphics{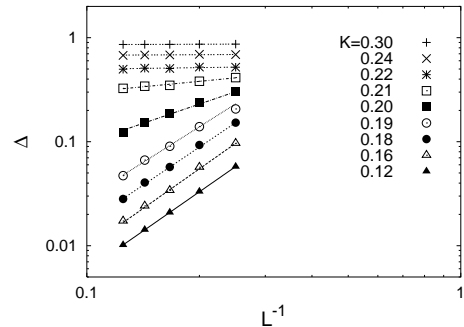}}}
\caption{
Log-log plots of $\Delta$ versus $L^{-1}$ for $d=5$ at various values of $K$.  
Lines are merely guides to eyes. 
}
\label{fig:5D_ph_L}
\end{figure}
For $d=5$, it looks evident that there exists an ordered (synchronized) phase 
extended to finite values of $K$. 
Similarly to the $d=4$ case, the log-log plots of $\Delta$ versus $L^{-1}$ 
are drawn in Fig.~\ref{fig:5D_ph_L}. 
For $K\lesssim 0.19$, we find the fully random phase: $\Delta\sim L^{-5/2}$. 
For $K\gtrsim 0.21$, on the other hand, $\Delta$, first decreasing slightly with $L$, 
eventually converges to a non-zero value. 
In fact, for $K\gtrsim 0.24$, this saturated value coincides perfectly well
with the linear-theory value: $\Delta=\exp[-\sigma/12\pi^2K^2]$. 
Note here that the linear theory breaks down for 
$K\lesssim K_0=\sqrt{\sigma/9}\approx 0.24$ and
the transition into the fully random phase apparently occurs a little later 
at $K_c\approx 0.20$.  It may be very interesting to understand this phase 
transition from the stability analysis in the weak coupling limit.

We next study the critical behavior near the synchronization transition.
%%where the order parameter $\Delta$ and the correlation length $\xi$ behave as
%%\begin{equation}
%%\Delta \sim (K-K_c)^{\beta} \ \ \mbox{and} \ \ \ \xi \sim |K-K_c|^{-\nu}.
%%\end{equation}
In a finite system, we assume the finite-size scaling relation
\begin{equation}
\Delta = L^{-\beta/\nu} f[(K-K_c)L^{1/\nu}],
\end{equation}
where the scaling function behaves $f(x)\sim x^\beta$ as $x\rightarrow +\infty$
and $f(x)\sim \mbox{constant}$ as $x\rightarrow 0$. 
At criticality, it leads to
\begin{equation}
\Delta(K_c,L)\sim L^{-\beta/\nu}.
\label{eq:betanu}
\end{equation}
%%In the unsynchronized phase ($K<K_c$),
%%we expect that $f\sim (-x)^{-\gamma/2}$ with $\gamma=d\nu-2\beta$
%%for $x\rightarrow -\infty$. This leads to 
%%$\Delta\sim L^{-d/2} (K_c-K)^{-\gamma/2}$, which is the characteristic
%%of the fully random phase.
To estimate efficiently the exponent $\beta/\nu$ and
the transition point $K_c$, we introduce the effective exponent
\begin{equation}
%\beta/\nu (L)=-\ln [\Delta(L{+}1)/\Delta(L)]/[\ln[(L{+}1)/L],
\beta/\nu (L)=-\ln [\Delta(L^{'})/\Delta(L)]/\ln(L^{'}/L),
\end{equation}
which is expected to approach zero, $\beta/\nu$, and $d/2$ 
for $K>K_c$, $K=K_c$, and $K<K_c$, respectively, as $L\rightarrow\infty$.

\begin{figure}
%\centering{\resizebox*{!}{4.5cm}{\includegraphics{5D_betanu_L_2.eps}}}
\centering{\resizebox*{!}{4.5cm}{\includegraphics{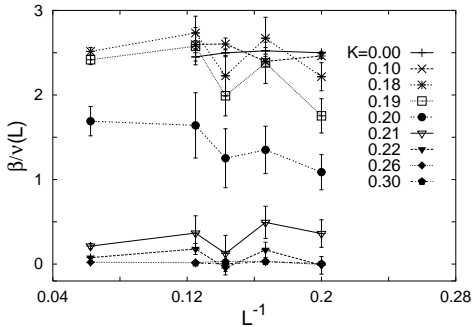}}}
%\centering{\resizebox*{!}{5.7cm}{\includegraphics{5D_betanu_slope.eps}}}
%\centering{\resizebox*{!}{5.7cm}{\includegraphics{../Figures/Datafiles/5D_betanu_crossing_1.eps}}}
%\centering{\resizebox*{!}{5.7cm}{\includegraphics{../Figures/Datafiles/5D_betanu_L_addlargeL.eps}}}
\caption{Effective exponent $\beta/\nu(L)$ versus $L^{-1}$ for $d=5$
at various values of $K$. 
%(b) The 
%slope gives us $(1-\beta)/\nu$ according to the Eq.~(\ref{eq:deriv}), 
%which yields $(1-\beta)/\nu= 0.6(2)$. 
%(c) Order parameter $\Delta$ 
%is plotted as $\Delta L^{\beta/\nu}$ with $\beta/\nu=1.5$ vs 
%$\exp(-K)$, displaying one unique crossing point at $K_c\approx 0.20$.
}
\label{fig:5D_beta_nu}
\end{figure}

The effective exponent for $d=5$, computed at various values of $K$, is plotted in 
Fig.~\ref{fig:5D_beta_nu}. 
The data for $K\le 0.19$ apparently converge to the weak-coupling value $5/2$, while
those for $K\ge 0.21$ converge to zero within statistical errors. 
Only the data at $K=0.20$ appear to converge to a nontrivial value. 
We thus estimate the critical coupling strength $K_c=0.200(5)$ and
%the exponent ratio $\beta/\nu=1.4(4)$. 
%the exponent ratio $\beta/\nu=1.4(3)$. 
the exponent ratio $\beta/\nu=1.6(3)$. 

\begin{figure}
%\centering{\resizebox*{!}{4.5cm}{\includegraphics{5D_scaling_1.eps}}}
%\centering{\resizebox*{!}{5.7cm}{\includegraphics{../Figures/Datafiles/5D_betanu_L_addlargeL.eps}}}
\centering{\resizebox*{!}{4.5cm}{\includegraphics{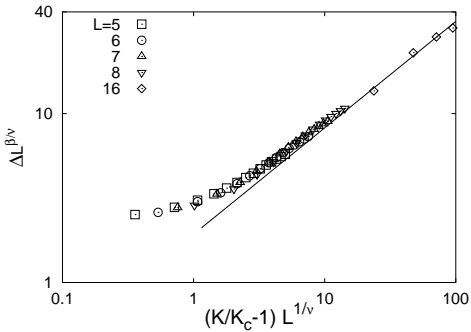}}}
\caption{Data collapse of $\Delta L^{\beta/\nu}$ against $(K/K_c -1)L^{1/\nu}$
in the log-log scale for various values of the system size and coupling strength. 
The best collapse is achieved with $\beta/\nu=1.4(3)$ and $\nu=0.45(10)$. 
The straight line has the slope 0.63, giving an estimation of $\beta$.}
\label{fig:5D_scaling}
\end{figure}

To check the finite-size scaling relation directly, we plot
$\Delta L^{\beta/\nu}$ versus $(K/K_c -1)L^{1/\nu}$ in Fig.~\ref{fig:5D_scaling}
and find that the data for various values of $L$ and $K$ are best collapsed
%to a curve with choices of $K_c=0.200(5)$, $\beta/\nu=1.4(2)$ and $\nu=0.45(10)$,
to a curve with choices of $K_c=0.200(5)$, $\beta/\nu=1.4(3)$ and $\nu=0.45(10)$,
which results in $\beta=0.63(20)$.  As expected, the resulting scaling function $f(x)$ 
converges to a constant for small $x$, and diverges as $x^\beta$ for large $x$
(see Fig. \ref{fig:5D_scaling}). 

We summarize our results for $d=5$:
\begin{equation}
%\beta/\nu =1.4(3),~~ \nu=0.45(10),~~ K_c =0.200(5). \nonumber
\beta/\nu =1.5(3),~~ \nu=0.45(10),~~ K_c =0.200(5). \nonumber
\end{equation}
Note the apparently substantial deviations from the MF values,
$\beta/\nu=1$ and $\nu=1/2$, although the latter may not be totally excluded. 
In view of the argument for the MF nature~\cite{ref:HHong}, 
these apparent deviations are rather unexpected and their origin is unclear at this stage. 
Similarly, we find for $d=6$:
%$\beta/\nu =1.0(5)$,  $\nu =??(?)$,  $K_c =0.160(5)$,
$\beta/\nu =1.0(3)$,  $\nu =0.45(10)$,  $K_c =0.158(5)$,
%%\begin{equation}
%%\beta/\nu =1.0(5),~~  \nu =??(?),~~  K_c =0.160(5), \nonumber
%%\end{equation}
which seem to be consistent with the MF values. 
%
%%Note that the obtained value $\beta/\nu=1.5(3)$ is much different
%%from $0.5$ for the mean-field system, implying that the 
%%five-dimensional system does not belong to the MF class. 
%%By means of the derivative of the scaling function at the critical
%%coupling strength $K_c$ we obtain $(1-\beta)/\nu$:
%%\begin{equation}
%%\ln \frac{d\Delta}{dK}\Bigg|_{K=K_c}=\frac{1-\beta}{\nu}\ln L+{\mbox const}.
%%\label{eq:deriv}
%%\end{equation} 
%%Figure~\ref{fig:5D_beta_nu} (b) exhibits that the slope is given by 
%%$(1-\beta)/\nu\approx 0.6(1)$.
%%From the two relations $\beta/\nu=1.5(3)$ and 
%%$(1-\beta)/\nu = 0.6(2)$, we obtain $\beta=0.7(1)(\simeq 3/4)$ and $\nu=0.5(1)$.
%%Figure~\ref{fig:5D_betanu_crossing} shows the plot of 
%%$\Delta L^{\beta/\nu}$ with $\beta/\nu=1.5$ as a function of $exp(-K)$, displaying 
%%one unique crossing point at $K_c \approx 0.20$.
%%Since we have measured the order parameter varying the increment 
%%$\delta K=0.01$, the critical coupling strength is given by  
%%$K_c=0.200(5)$. 

In summary, we have explored the phase synchronization phenomena 
in the system of locally-coupled oscillators with scattered intrinsic 
frequencies on $d$-dimensional lattices.  
A linear analysis shows that the strong coupling regime can be described by the
EW surface growth equation with columnar disorder for $d\ge3$. 
It has been shown analytically that the system is always desynchronized up to $d=4$,
while numerical integration for $d\ge 5$ has demonstrated the emergence of 
the synchronized (ordered) phase via a continuous transition from the 
desynchronized phase.  The lower critical dimension for phase synchronization
is thus given by $d_{l}=4$, but the critical behavior explored for $d=5$ and 6 
does not give a conclusive result for the upper critical dimension. 

We thank P. Grassberger for useful discussions at the initial stage of this 
work.  This work was supported in part by Grant No.~2000-2-11200-002-3
from the Basic Research Program of KOSEF (H.P.) and 
by the Ministry of Education through the BK21 Program (M.Y.C.).


\begin{thebibliography}{20}
\bibitem{explanation1} The coupling constant $K$ and the variance $2\sigma$ may be
renormalized through a coarse graining procedure in taking the continuum limit.
However, the renormalizing factor may be absorbed into the time scale via proper
rescaling of time.
\bibitem{explanation2} Numerical values of $K_0$ at which the linear theory begins to
%break down are consistent with the analytical estimates by setting $G^2(K_0)=d/2$.
break down are consistent with the analytical estimates by setting $G^2(K_0)=1/2$.
\end{thebibliography}
\end{document}